\documentclass[12pt]{article}
\begin{document}

\title{\bf Homotheties of Cylindrically Symmetric Static Manifolds
and their Global Extension}
\author{Asghar Qadir$^\dag$\thanks{Senior Associate, Abdus Salam International Centre
for Theoretical Physics, Trieste, Italy.}~, M. Sharif$^\ddag$ and M. Ziad$^\dag$\\
$\dag$ Department of Mathematics,\\ Quaid-i-Azam University,
Islamabad, Pakistan\\
$\ddag$ Department of Mathematics, Punjab University\\ Lahore,
Pakistan.}

\date{}
\maketitle

\begin{abstract}
Cylindrically symmetric static manifolds are classified according to
their homotheties and metrics. In each case the homothety vector
fields and the corresponding metrics are obtained explicitly by
solving the homothety equations. It turns out that these metrics
admit homothety groups $H_m$, where $m=4,5,7,11$. This
classification is then used to identify the cylindrically symmetric
static spaces admitting the local homotheties, which are globally
prohibited due to their topological construction. Einstein's field
equations are then used to identify the physical nature of the
spaces thus obtained.
\end{abstract}
\textbf{PACS numbers:} 0240, 0490
\vspace{.5cm}

Homotheties [1] of metrics are defined to be the vector fields
$\textbf{H}$, along which the metric tensor $(g)$ remains invariant
upto constant scale factor $(\phi).$ This amounts to saying that
\begin{equation}
L_{H}g=2\phi g,
\end{equation}
where $\textbf{H}$ reduces to an isometry, or a Killing vector field
(KV) if $\phi=0$. For $\textbf{H}$ to be a proper homothety vector
field, $\phi \neq0$. In this paper proper homotheties are used to
distinguish the static cylindrically symmetric manifolds which admit
$SO(2)$ locally rather than globally and identify the spaces which
have topological defects.

In a local coordinate patch, equation (1) is a system of ten partial
differential equations (PDEs) involving product terms of first-order
partial derivatives of the four components
$H^{a}(x^{b}),~(a,b,...=0,1,2,3)$ of the vector field
$\textbf{H}=H^a\partial/\partial x^a$ and of the ten metric
coefficients, all depending in general upon four variables. Due to
their physical significance [2], various authors have found
self-similar solutions of Einstein's equations by imposing some
constraints on the geometry or on the matter content of the
spacetimes [3]. Hall and Steele [4] investigated the Segr$\acute{e}$
and Petrov types of spaces that admit proper homothety groups. They
classified all such gravitational fields for homothety groups,
$H_{m}$, with $m\geq 6$, whereas for $m\leq5$, some general remarks
were given. However, they did not provide the metrics along with the
generators of $H_m$ explicitly.

An alternative approach [5], which was adopted earlier to obtain
complete classifications of manifolds admitting some minimal
isometry groups, was further enhanced to obtain the homotheties of
spherically symmetric and plane-symmetric static space [6].
Einstein's field equations were then used as the defining equations
for the stress-energy tensor. Here the same method is used to obtain
the homotheties and the corresponding metrics for cylindrically
symmetric static spacetimes (CSSS), written in generalized
cylindrical coordinates $(x^0=ct,~ x^1=\rho,~x^2=\theta,~x^3=z)$ as
[7]
\begin{equation}
ds^2=\exp[\nu(\rho)]dt^2-d\rho^2-\exp[\lambda(\rho)]a^2d\theta^2-\exp[\mu(\rho)]dz^2,
\end{equation}
where '$a$' has units of length. Thus the metric admits the group of
motions $SO(2)\bigotimes R^2$, where
$SO(2)=(\partial/\partial\theta)$, one $R=(\partial/\partial{z})$
and the other $R=(\partial/\partial_{t})$, as the minimal isometry
group. The usual cylindrically symmetric spaces have an axis
definable by the limit $ \rho \rightarrow 0$. This is analogous to
the usual spherically symmetric spaces having an origin as
$r\rightarrow 0$. However, there are additional spaces which are
classified as spherically symmetric, in that the manifolds have
spacelike 2-sections of constant curvature. These are the
Bertotti-Robinson and similar metrics. They do not have any origin
in the above sense. Correspondingly, for the cylindrically symmetric
spaces, there are those which admit the required group
$SO(2)\bigotimes R^2$ but do not have an axis. We will call these
'cylindrical Bertotti-Robinson metrics'. For the usual cylindrically
symmetric manifolds, the difference between an $SO(2)$ and an $R$ is
easily seen. For the cylindrical Bertotti-Robinson metrics, it is
not \textit{a priori} clear how to distinguish between them. Using
homotheties we are able to distinguish them.

The metric (2) with $\nu=\lambda=\mu=0,~a\theta$ replaced by $y$ and
$\rho$ by $x$ can easily be recognized as the Minkowski space which
was called the 'wrapped Minkowski space' in the classification of
CSSS according to their isometries [8]. In fact, the Minkowski space
written in cylindrical coordinates acquired special interest with
its application\footnote{The simplest example of a topological field
theory constructed on a 'wrapped up' Minkowski space was presented
by R Rajaraman at the \emph{2nd BCSPIN Summer School in Physics}
held at Kathmandu, Nepal in 1991. However, it does not appear either
in his write-up of the lectures in [9], or in the standard
reference.} to 'topological field theories'. This metric could
alternatively be identified as a CSSS, having a line singularity at
$\theta=0, 2\pi$, with a topological defect.

Equations (1) written in local coordinates,
\begin{equation}
(H_{ab}):~g_{ab,c}H^{c}+g_{ac}H^c_{,b}+g_{bc}H^c_{,a}=2\phi g_{ab,}
\end{equation}
are solved for the metric (2) to obtain the required classification.
Comparing $(H_{i2})_{,3}$ and $(H_{i3})_{,2}~(i=0,1)$, using
$(H_{23})$ and $(H_{13})$, one obtains
\begin{equation}
H^0_{,23}=H^2_{,03}=H^3_{,02}=0_{,}
\end{equation}
\begin{equation}
H^1_{,23}=e^{\mu}[(\mu'-\lambda')H^3_{,2}+H^3_{,21}]
\end{equation}
and
\begin{equation}
(e^{(\mu-\lambda)/2}H^3_{,2})_{,1}=0,
\end{equation}
where ',' denotes differentiation relative to $\rho$. Combining
equations (4) and (6), one obtains
\begin{equation}
(e^{(\mu-\lambda)/2}H^3_{,2})_{,i}=0,
\end{equation}
which gives
\begin{equation}
H^3_{,2}=e^{(\mu-\lambda)/2} f(\theta, z).
\end{equation}
Thus from $(H_{23})$
\begin{equation} H^2_{,3}=-e^{(\mu-\lambda)/2}
f(\theta, z),
\end{equation}
where $f(\theta{, z})$ is a function of integration. Now, using
$(H_{13})_{,2},~(H_{22})_{,32},~(H_{33})_{,32}$ and equations (8)
and (9), yields
\begin{equation}
f(\theta,z)_{,22}+\alpha f(\theta,z)=0,
\end{equation}
and
\begin{equation}
f(\theta,z)_{,33}-\beta f(\theta,z)=0,
\end{equation}
where $\alpha$ and $\beta$ are separation constants given by
\begin{equation}
\lambda'(\lambda'-\mu')e^{\lambda}=4 \alpha,
\end{equation}
and
\begin{equation}
\mu'(\lambda'-\mu')e^{\mu}=4 \beta.
\end{equation}
Considering all possible combinations of $\alpha$ and $\beta$ and
using the rest of equation $(H_{ab})$, one obtains a complete
solution of equation (3), giving $H^a$ and the corresponding metric
components. These results are listed according to their homothety
groups, $H_m$, as follows.

There are three metrics admitting four homotheties:\\
(a)
\begin{eqnarray}
\nu&=&2\alpha\ln(\rho/\rho_{0}),\quad\lambda=2\ln(\rho/a),\quad\mu=0,\quad(\alpha
\neq 0,1) \nonumber\\
H^0&=&(1-\alpha)\phi t+C_{0},\quad H_1=\rho\phi,\quad H^2=C_2,\quad
H^3=\phi z+C_{3,}
\end{eqnarray}
of Petrov type I [8] and Segr$\acute{e}$ type [1,111] [8];\\
(b)
\begin{eqnarray}
\nu&=&2\alpha\ln(\rho/\rho_{0}),\quad\lambda=\mu=2\beta
\ln(\rho/\rho_0),\quad(\alpha\neq\beta~
and~ \alpha, \beta\neq 0,1) \nonumber \\
H^0&=&(1-\alpha)\phi t+C_{0},\quad H_1=\rho\phi,\nonumber\\
H^2&=&(1-\beta)\phi\theta+C_2,\quad H^3=(1-\beta)\phi z+C_{3,}
\end{eqnarray}
of Petrov type D and Segr$\acute{e}$ type[1,(11)1];\\
(c)
\begin{eqnarray}
\nu&=&2\alpha\ln(\rho/\rho_{0}),\quad\lambda=0,\quad\mu=2\ln(\rho/\rho_{0}),\nonumber\\
H^0&=&(1-\alpha)\phi,\quad H^1=\rho\phi,\quad
H^2=\phi\theta+C_2,\quad H^3=C_3,
\end{eqnarray}
of Petrov type I and Segr$\acute{e}$ type [1, 111].\\
There are three metrics admitting five homotheties, which are:\\
(a)
\begin{eqnarray}
\nu&=&2\ln(\rho/\rho_{0}),\quad\lambda=2\ln(\rho/a),\quad\mu=0,\nonumber\\
H^0&=&C_{1}\theta+C_{0},\quad H^1=\rho\phi,\nonumber\\
H^2&=&(C_1/\rho^{2}_{0})t+C_{3},\quad H^3=\phi z+C_{3},
\end{eqnarray}
of Petrov type D, Segr$\acute{e}$ type [1, 111];\\
(b)
\begin{eqnarray}
\nu&=&\mu=2\ln(\rho/\rho_{0}),\quad\lambda=0,\nonumber \\
H^0&=&zC_1+C_{0},\quad H^1=\rho \phi,\nonumber\\
H^2&=&\phi \theta + C_{3},\quad H^3=C_{1}t+C_{3},
\end{eqnarray}
of Petrov type D, Segr$\acute{e}$ type [1,111],\\
(c)
\begin{eqnarray}
\nu&=&0,\quad\lambda=2\ln(\rho/a)=\mu,  \nonumber\\
H^0&=&\phi t+C_{0},\quad H^1=\rho\phi, \nonumber\\
H^2&=&(-z/a)C_4+C_2,\quad H^3=a\theta C_4+C_3,
\end{eqnarray}
of Petrov type D, Segr$\acute{e}$ type [(1,1)(11)].\\
There is only one metric admitting seven homotheties:
\begin{eqnarray}
\nu&=&\lambda=\mu=2\alpha \ln(\rho/\rho_0),\quad(\alpha\neq 0,1)
\nonumber\\
H^0&=&C_1z+\alpha\theta C_3+(1-\alpha)\phi t+C_0,\nonumber\\
H^1&=&\rho\phi, \nonumber\\
H^2&=&(-z/a)C_4+(1-\alpha)\phi \theta+(t/a)C_3+C_6,\nonumber\\
H^3&=&a\theta  C_4+(1-\alpha)\phi z+C_1t+C_2,
\end{eqnarray}
of Petrov type D, Segr$\acute{e}$ type [1,(11)1].

There are three metrics admitting 11 homotheties, other than proper
Minkowski space which admits $\textbf{H}=\phi(t\partial/\partial
t+\rho\partial/\partial\rho+z\partial/\partial z)$ as a proper
homothety vector. They are:
\begin{eqnarray}
(a)\quad\nu&=&\lambda=0,\quad\mu=2\ln(\rho/\rho_0)\quad \rm{with}
\nonumber\\
\textbf{H}&=&\phi(t\partial/\partial t+ \rho \partial/\partial \rho
+ \theta
\partial/\partial \theta)+ z
\partial/\partial z),\\
(b)\quad \nu&=&\lambda=\mu=0,\quad \rm{with} \quad
\textbf{H}=\phi(t\partial/\partial t+ \rho
\partial/\partial \rho + \theta \partial/\partial \theta+z \partial/\partial z),\\
(c)\quad \nu&=&2\ln(\rho/\rho_0),\quad\lambda=\mu=0\quad \rm{with}
\nonumber\\
\textbf{H}&=&\phi(\rho \partial/\partial \rho+\theta
\partial/\partial \theta+z\partial/\partial z).
\end{eqnarray}

In equations (14)-(23), there are two types of solutions: one in
which the $H^2$ component of the proper homothety vector field is
zero; and the other in which it is non-zero and involves $\theta$.
This shows that in the first type of solution, the homotheties can
be extended globally, whereas in the latter this is not possible due
to a line singularity at $\theta=0,2 \pi$. This is how one can
identify the topologically constructed spaces, even if they are
non-flat.

As a result of this classification, we see that there exist cases of
four, five, seven and 11 homotheties. The metrics given by equations
(21)-(23); equation (20); equation (18) and equations (15) and (16)
admitting 11, seven, five and four homotheties, respectively, do not
appear in the classification of Hall and Steele [4] due to their
topological construction, as these admit homotheties which cannot be
extended globally. However, the spaces other than the proper
Minkowski space given by equations (14) and (17), where the
homothety vector fields could be globally extended, match their
solution. The metric given by equation (19) can be considered to be
that of equation (17) written in different coordinates.

The metric given by equation (15) represents a perfect fluid for
$\beta=\alpha(\alpha-1)/(\alpha+1)$, whereas it represents a
non-null electromagnetic field for $\beta=\alpha+1$. Both conditions
are satisfied when $\alpha=-\frac{1}{3}$ and $\beta=\frac{2}{3}$.
The metric given by equation (19) represents a non-null
electromagnetic field. The metric given by equation (20) represents
a tachyonic fluid and could be re-interpreted as an anisotropic
fluid with an appropriately chosen cosmological constant. All other
metrics are non-physical. Of course, any non-physical metric could
be re-interpreted as a physical metric with an appropriately chosen
cosmological constant. However, these metrics are not of particular
interest.

\vspace{0.5cm}

\textbf{Acknowledgements}

\vspace{0.5cm}

We are extremely grateful to the referees for their highly useful
comments. Two of the authors (AQ and MZ) would like to thank
Professor M. Virasoro and the Abdus Salam ICTP for hospitality while
this work was being done. One of the authors (MS) is grateful to the
Third World Academy of Sciences Associateship Scheme for support to
travel to Brazil and to the State University of Campinas for local
hospitality, while the work was being done. This work was partially
supported under Pakistan Science Foundation Project
PSF/RES-C-QU/MATHS(21).

\vspace{0.5cm}

\textbf{References}

\vspace{0.5cm}

\begin{description}

\item{[1]} Katzin G.H., Levine J. and Davis W.R. J. Math. Phys.
\textbf{10}(1969)617;\\
Katzin G.H., Levine J. and Davis W.R. J. Math. Phys.
\textbf{11}(1970)1518.

\item{[2]} Cahill M.E. and Taub A.H. Commun. Math. Phys.
\textbf{21}(1971)1;\\
Taub A.H. \emph{General Relativity} ed L O' Raifertaigh (Oxford:
Oxford University Press, 1972)p 135-50;\\
Winicour J. \emph{Lecture Notes in Physics vol 14} (Berlin:
Springer,
1972)p 145;\\
Eardley D.M. Commun, Math, Phys. \textbf{37}(1974)287;\\
Mclntosh C.B.G. Phys. Lett. \textbf{A50}(1975)429;\\
Mclntosh C.B.G. Gen. Rel. Grav. \textbf{7}(1976)199;\\
Mclntosh C.B.G. Gen. Rel. Grav. \textbf{7}(1976)215.

\item{[3]} Yano K. J. Indian Math. Soc. (NS) \textbf{15}(1951)105;\\
Ehlers J. and Kundt W. \emph{Gravitation: An Introduction to Current
Research} ed L. Witten (New York: Wiley 1962)p 80;\\
Collinson C.D. and French D.C. J. Math. Phys. \textbf{8}(1967)701;\\
Godfrey B.B. Gen. Rel. Grav. \textbf{3}(1972)3;\\
Thompson A. Phys. Rev. Lett. \textbf{349}(1975)507;\\
Halford W.D. and Kerr R.P. J. Math. Phys. \textbf{21}(1980)120;\\
Halford W.D. J. Math. Phys. \textbf{21}(1980)129;\\
Hall G.S. Gen. Rel. Grav. \textbf{20}(1988)671.

\item{[4]} Hall G.S. and Steele J.D. Gen. Rel. Grav.
\textbf{22}(1990)457.

\item{[5]} Bokhari A.H. and Qadir A. J. Math. Phys.
\textbf{28}(1987)1019;\\
Bokhari A.H. and Qadir A. J. Math. Phys. \textbf{31}(1990)1463;\\
Qadir A. and Ziad M. J. Math. Phys. \textbf{29}(1988)2473;\\
Qadir A. and Ziad M. J. Math. Phys. \textbf{31}(1989)254;\\
Qadir A. and Ziad M. Nuovo Cimento \textbf{B110}(1995)317;\\
Qadir A. and Ziad M. \emph{Proc. 6th Marcel Grossmann Meeting} ed T.
Nakamura and H. Sato (Singapore: World Scientific, 1993)p 1115-8.

\item{[6]} Ahmad D. and Ziad M. J. Math. Phys. \textbf{38}(1997)2547;\\
Ziad M. \emph{The classification of static plane symmetric
spacetimes} Nuovo Cimento \textbf{B114} (to appear).

\item{[7]} Kramers D., Stephani H., MacCallum M.A.H. and Herlt E. \emph{Exact
Solutions of Einsterin's Field Equations} (Berlin: Deutscher, 1980).

\item{[8]} Qadir A. and Ziad M. Nuovo Cimento \textbf{B110}(1995)277.

\item{[9]} Rajaraman R. \emph{Current Topics in Condensed matter and
Particle Physics} ed J.C. Pati, Q. Shafi and Y. Lu (Singapore: World
Scientific, 1993).
\end{description}
\end{document}